# Potential. Solution of Poisson's Equation, Equations of Continuity and Elasticity

### Alexander Ivanchin

*The modern theory of the potential does not give a solution of Poisson's equation. In the present work its solution has been found via generalized functions and a nonpotential solution of the continuity equation has been obtained. The method is demonstrated by the solution of elasticity equations using the example of a crack in the infinite specimen and a surface crack. Their elastic energies have been calculated. In has been shown that there is no critical condition for a crack in the infinite specimen and the crack always closes. Only the surface crack possesses the critical condition.*

**Introduction**

In physics the method of the potential is used to solve differential equations. The present-day theory of the potential does not offer a solution of Poisson's equation. In the present work the method of the potential has been examined on the basis of generalized functions [1], a solution of Poisson's equation has been obtained and a nonpotential solution of the continuity equation has been derived. At the beginning of the article some data are presented from the theory of generalized functions important for the description of the theory of the potential.

**1. The method of the potential**

The following symbols are used

$\nabla \bullet = \left\{ \dfrac{\partial}{\partial x}, \dfrac{\partial}{\partial y}, \dfrac{\partial}{\partial z} \right\} \bullet$ - for the gradient,

$\nabla = \left( \dfrac{\partial}{\partial x} + \dfrac{\partial}{\partial y} + \dfrac{\partial}{\partial z} \right)$ - for the divergence,

$\nabla \times = \left\{ \dfrac{\partial}{\partial x}, \dfrac{\partial}{\partial y}, \dfrac{\partial}{\partial z} \right\} \times$ - for the rotor,

$\Delta = \left( \dfrac{\partial^2}{\partial x^2} + \dfrac{\partial^2}{\partial y^2} + \dfrac{\partial^2}{\partial z^2} \right)$ - for the Laplace operator.

The fundamental solution of the equation for the Laplace operator is the solution of the equation

$$\Delta \Gamma(x) = \delta(x) \qquad (1.1)$$

Here $\delta(x)$ is the Dirac function. The Poisson equation is

$$\Delta \Phi = 2\pi F(\mathbf{r}) \qquad (1.2)$$

$F(\mathbf{r})$ is the known function. According to the present-day knowledge, the solution (1.2) is the function $\Phi = \Gamma * F$.

The symbol $*$ stands for convolution. Convolution of the two functions $f_1$ and $f_2$ is

$$f_1 * f_2 = \int f_1(t) f_2(x-t) dt = \int f_1(x-t) f_2(t) dt$$

$x$ and $t \in R^n$. Integration is performed over the whole space. The convolution has the following differential property

$$\left(\frac{\partial^n}{\partial x^n} f_1\right) * f_2 = \frac{\partial^n}{\partial x^n}(f_1 * f_2) = f_1 * \left(\frac{\partial^n}{\partial x^n} f_2\right) \tag{1.3}$$

In addition to the Dirac $\delta$-function determined at a point, here its generalization is used $\mu(S)\delta_S$, i.e. the function determined on the surface $S$ in the following way [1]

$$\int_V F(x)\mu(S)\delta_S \, dV = \int_S F(x)\mu(S)dS \tag{1.4}$$

The generalized function is the linear functional determined from a set of finite functions. The supplier of the generalized function is the set in which the generalized function is not zero. For the Dirac $\delta$- function the supplier is a point and for $\mu(S)\delta_S$ - the surface $S$.

The fundamental solution (1.1) for the two-dimensional case is

$$\Gamma(x, y) = \frac{1}{2\pi} \ln\sqrt{x^2 + y^2} \tag{1.5}$$

and for the three-dimensional case

$$\Gamma(x, y, z) = \frac{1}{4\pi}\frac{1}{r} \tag{1.6}$$

Further we shall consider the two-dimensional case. Extension to the three-dimensional case is not a problem.

According to the present-day knowledge, a potential, that is a solution of the equation (1.2) for the two-dimensional case is

$$\Phi(x, y) = 2\pi \int_{-\infty}^{\infty} \int_{-\infty}^{\infty} F(t_1, t_2)\ln\sqrt{(x-t_1)^2 + (y-t_2)^2}\, dt_1 dt_2 \tag{1.7}$$

The gradient of $\Phi(x, y)$ is the vector $\hat{U}$

$$\hat{U}_x = \frac{\partial \Phi}{\partial x} = \frac{1}{2\pi} \int_{-\infty}^{\infty} \int_{-\infty}^{\infty} F(t_1, t_2)\frac{x - t_1}{(x-t_1)^2 + (y-t_2)^2} dt_1 dt_2$$

$$\hat{U}_y = \frac{\partial \Phi}{\partial y} = \frac{1}{2\pi} \int_{-\infty}^{\infty} \int_{-\infty}^{\infty} F(t_1, t_2)\frac{y - t_1}{(x-t_1)^2 + (y-t_2)^2} dt_1 dt_2 \tag{1.8}$$

So we derive

$$\Delta\Phi = 2\pi \iint F(t_1, t_2)\Delta\ln\sqrt{(x-t_1)^2 + (y-t_2)^2}\, dt_1 dt_2 =$$
$$= \iint F(t_1, t_2)\delta(x-t_1)\delta(y-t_2)dt_1 dt_2 = F(x, y) \tag{1.9}$$

The second derivatives of (1.7) are written as

$$\frac{\partial^2 \Phi}{\partial x^2} = -\frac{\partial^2 \Phi}{\partial y^2} = \frac{1}{2\pi}\iint F(t_1, t_2)\frac{-(x-t_1)^2 + (y-t_2)^2}{[(x-t_1)^2 + (y-t_2)^2]^2} dt_1 dt_2 \tag{1.10}$$

that is (1.7) is really a solution of the Laplace equation $\Delta\Phi = 0$ rather than the Poisson equation (1.2).

When explaining this paradox, it is usually pointed out that at $x = t_1$ and $y = t_2$ in (1.7) the numerator and the denominator become zero, with the zero of the denominator being of a higher order than that of the numerator. What is meant by that is not usually explained. Actually, one should be careful or rather accurate when differentiating than integrating. If one deals with generalized functions, then differentiation should be understood in the generalized sense. For the classical derivatives $\Delta\ln r = 0$ across the whole plane except for the origin of coordinates where $\ln r$ is not determined. The relation $\Delta\ln r = \frac{1}{2\pi}\delta$ is derived during differentiation in the generalized sense. Further classical derivatives will be designated by the Latin letters $\partial/\partial x$ and the generalized ones by the capital $D_x$, with the lower index indicating the differentiation variable and $\Delta_D$ standing for the Laplace operator with the generalized derivatives $\Delta_D$.

In the classical sense, along with (1.5) the solution of the Laplace equation is also the function

$$\varphi(x, y) = \frac{1}{2\pi} \arctan\left(\frac{y}{x}\right) \tag{1.11}$$

which, unlike (1.5) is discontinuous. In tracing the boundary around the origin of coordinates it increases by 1 and has a jump equal to 1 along the positive part of the abscissa. Therefore, the generalized derivative of (1.11) will be written as

$$D_x \varphi(x, y) = \frac{\partial}{\partial x}\left[\frac{1}{2\pi} \arctan\left(\frac{y}{x}\right)\right]$$

$$D_y \varphi(x, y) = \frac{\partial}{\partial y}\left[\frac{1}{2\pi} \arctan\left(\frac{y}{x}\right)\right] + \delta_{+x}$$

and its Laplacian in the generalized sense is

$$\Delta_D \varphi = D_y \delta_{+x}$$

Here $\delta_{+x}$ is the Dirac function having the supplier as the positive part of the abscissa axis [1]. According to (1.4), for the finite function $\phi$ we have

$$\iint \phi(x, y) \delta_{+x}(x, y) dx dy = \int_0^\infty \phi(x, 0) dx = \iint \phi(x, y) \delta(y) \Theta(x) dx dy$$

Hence we obtain

$$\delta_{+x}(x, y) = \delta(y) \Theta(x) \tag{1.12}$$

Here

$$\Theta(x) = \begin{cases} 1, & x > 0 \\ 0, & x < 0 \end{cases}$$

is the Heaviside function. One arrives at the relation

$$D_x \Theta(x) = \delta(x)$$

The derivative of $F$ is determined in this way

$$\iint F(x, y) \frac{\partial}{\partial x} \phi(x, y) dx dy = -\iint D_x F(x, y) \phi(x, y) dx dy \tag{1.13}$$

For classical derivatives the following equalities take place

$$\frac{\partial}{\partial x} \ln \sqrt{x^2 + y^2} = \frac{\partial}{\partial y} \arctan\left(\frac{y}{x}\right)$$

$$\frac{\partial}{\partial y} \ln \sqrt{x^2 + y^2} = -\frac{\partial}{\partial x} \arctan\left(\frac{y}{x}\right) \tag{1.14}$$

From (1.13) and (1.14) it follows that the above equalities hold for the generalized derivatives

$$D_x \ln \sqrt{x^2 + y^2} = D_y \arctan\left(\frac{y}{x}\right)$$

$$D_y \ln \sqrt{x^2 + y^2} = -D_x \arctan\left(\frac{y}{x}\right) \tag{1.15}$$

In [1] the following equality is proved

$$D_{i_i} F = \frac{\partial}{\partial x_i} F + [F]_S n_{i_i}$$

Here $S$ is the surface on which function $F$ has a jump, which is denoted by $[F]_S$, $\mathbf{n} = \lfloor n_{i_i} \rfloor$ is normal to the surface $S$. The $\arctan(y/x)$ is discontinuous on the positive side of the abscissa axis and, therefore, $\mathbf{n} = [0, -1]$. The value of the jump is $[\arctan(y/x)]_{+x} = 2\pi$, then from (1.13)–(1.15) it follows that

$$D_x \ln\sqrt{x^2+y^2} = D_y \arctan\left(\frac{y}{x}\right) = \frac{\partial}{\partial x}\ln\sqrt{x^2+y^2} + 2\pi\delta_{+x}$$

$$D_y \ln\sqrt{x^2+y^2} = \frac{\partial}{\partial y}\ln\sqrt{x^2+y^2}$$

(1.16)

The equations in (1.16) solve the contradiction in (1.9) and (1.10). Substituting (1.16) into (1.8) instead of the classical derivatives we obtain the solution of the Poisson equation (1.2) in the form

$$\mathbf{U} = C\hat{\mathbf{U}} + \tilde{\mathbf{U}}$$

(1.17)

Here $C$ is the arbitrary constant and the vector $\tilde{\mathbf{U}}$ by (1.16) is written as

$$\tilde{U}_x = \hat{U}_x = \iint F(t_1,t_2)\delta_{+x}(x-t_1, y-t_2)dt_1 dt_2$$

$$\tilde{U}_y = 0$$

(1.18)

The derivative of $\tilde{U}_x$ with respect to $x$ according to (1.12) looks like

$$\frac{\partial}{\partial x}\tilde{U}_x = \iint F(t_1,t_2) D_x \Theta(x-t_1)\delta(y-t_2)\, dt_1 dt_2 = \iint F(t_1,t_2)\, \delta(x-t_1)\delta(y-t_2)\, dt_1 dt_2 = F(x,y)$$

.

If the generalized function $\omega$ has a point as a supplier, it is presented, in the general case, like this [1]

$$\omega = \sum_{k=0}^{\infty} C_k D^{(k)}\delta(r)$$

(1.19)

Here $C_k$ is the arbitrary constants, and $D^{(k)}$ is the generalized derivative of the $k$-order. If the supplier of the generalized function is not a point but a line or a surface, then the formula (1.19) is generalized for this case

$$\omega_S = \sum_{k=0}^{\infty} C_k D^{(k)}\mu(S)\delta_S$$

(1.20)

It is sufficient to find a solution of the differential equation with $\omega = \delta$ and derive from it a general solution (1.19) and (1.20). From the general solution using some additional conditions one can obtain a solution that is implemented in physics. It is impossible to determine the form of the additional conditions in a general case, as they are chosen specifically for a particular problem.

In physics when solving the equation

$$\nabla \bullet \mathbf{U} = 0$$

(1.21)

the following technique is used. It is said that suppose the unknown function $\mathbf{U}$ is potential, that is, it is shown in the form

$$\mathbf{U} = \nabla\Psi$$

(1.22)

then (1.21) is reduced to the Laplace equation

$$\Delta\Psi = 0$$

whose solution is (1.5) or (1.6). The supposition on potentiality does not follow from the physical laws, but it is made to solve the problem in some way. The obtained solution is considered unique. It is a mistake. Let us consider here another nonpotential solution (1.22). Substituting the nonpotential vector $\tilde{\mathbf{U}}$ from (1.18) into (1.21.) we arrive at

$$\nabla \bullet \tilde{\mathbf{U}} = F$$

(1.23)

Let us find a solution of the Poisson equation
$$\Delta \Psi = F$$
Then the vector $\tilde{\mathbf{U}} - \nabla \Psi$ will be a nonpotential solution of the equation (1.21) meeting the same boundary conditions. Examples of obtaining such solutions for particular physical problems are given below as well as in [2,3].

Let $F(x)$ be primitive of $f(x)$, then we get the following equality

$$\int_a^b f(t)\operatorname{sign}(t)dt = F(a) + F(b) - 2F(0) \tag{1.24}$$

The theory of elasticity is chosen specially to illustrate the suggested method, because the equations of the elastic equilibrium are not Poisson's equations in a pure form. However, it is possible to reduce the above problem to the solution of the Poisson equation. Besides, the theory of elasticity allows one to illustrate the application of the generalized functions different from the point $\delta$-function.

**2. Solution of the Equation of the Elasticity Theory by the Method of the Potential**

For an elasto-isotropic body the stress tensor $\sigma_{ij}$ and the deformation tensor $\varepsilon_{ij}$ are related via the linear dependence [8]

$$\sigma_{ij} = 2G\left[\varepsilon_{ij} + \left(\frac{1}{2a} - 1\right)\delta_{ij}\,\varepsilon_{ij}\right] \tag{2.1}$$

Here $G$ is the shear module, $v$ is the Poisson coefficient,

$$a = \frac{1-2v}{2(1-v)} \tag{2.2}$$

$\delta_{ij} = \begin{cases} 1, & i = j \\ 0, & i \neq j \end{cases}$ is the Kronecker symbol, $\mathbf{U}$ is the elastic displacement vector. The deformation tensor is

$$\varepsilon_{ij} = \frac{1}{2}\left(\frac{\partial U_i}{\partial x_j} + \frac{\partial U_j}{\partial x_i}\right) \tag{2.3}$$

The equation of elastic equilibrium is written as [8]:
$$\nabla \nabla \bullet \mathbf{U} - a \nabla \times \nabla \times \mathbf{U} = 0 \tag{2.4}$$

The sources of displacement are designated by vector $\mathbf{X}$ and include plastic deformation, thermal expansion and internal ruptures: cracks, pores, etc. Then the elastic displacement is expressed in the form
$$\mathbf{U} = \mathbf{T} - \mathbf{X} \tag{2.5}$$
Here $\mathbf{T}$ is the total displacement. Substituting (2.5) into (2.4) we obtain the equation
$$\nabla_D \nabla_D \bullet \mathbf{T} - a \nabla_D \bullet \nabla_D \bullet \mathbf{T} = \nabla_D \nabla_D \bullet \mathbf{X} - a \nabla_D \bullet \nabla_D \bullet \mathbf{X} \tag{2.6}$$
The sources of displacement can be concentrated, that is non-zero, at points, lines, surfaces. The generalized functions are needed to describe them [1]. Outside the carrier of the generalized function the equality $\mathbf{U} = \mathbf{T}$ takes place, that is, outside the carrier (2.6) coincides with (2.4). Applying the operation of divergence and rotor to (2.6) we obtain a system of equations
$$\Delta_D(\nabla_D \bullet \mathbf{T}) = \Delta_D(\nabla_D \bullet \mathbf{X}) \tag{2.7}$$
$$\Delta_D(\nabla_D \times \mathbf{T}) = \Delta_D(\nabla_D \times \mathbf{X})$$
Here the differential equalities are used
$$\nabla_D \times \nabla_D \times = \nabla_D(\nabla_D \bullet) - \Delta_D$$
$$\nabla_D(\nabla_D \times) = 0$$

If we remove the similar differential operators in (2.7) and writes down $\mathbf{T} = \mathbf{X}$, then we derive a trivial solution $\mathbf{U} \equiv 0$, which is not of interest. (2.7) are Poisson's equations with respect to $\nabla_D \bullet \mathbf{T}$ and $\nabla_D \times \mathbf{X}$. Let us consider the first equation (2.7), and the second is solved in the same way. According to (1.17), the solution (2.7) is

$$\nabla_D \bullet \mathbf{T} = C(\nabla_D \bullet \nabla_D \nabla_D \bullet \mathbf{X}) * \Gamma + \nabla_D \bullet \tilde{\mathbf{U}} = C\nabla_D \bullet \mathbf{X} + \nabla_D \bullet \tilde{\mathbf{U}} \tag{2.8}$$

By (1.3) we get
$$(\nabla_D \bullet \nabla_D \nabla_D \bullet \mathbf{X}) * \Gamma = (\Delta_D \nabla_D \bullet \mathbf{X}) * \Gamma = (\nabla_D \bullet \mathbf{X}) * (\Delta_D \Gamma) = (\nabla_D \bullet \mathbf{X}) * (\delta(x)\delta(y)) = \nabla_D \bullet \mathbf{X}$$

According to (1.8)
$$\Delta \nabla \bullet \tilde{\mathbf{U}} = 0 \tag{2.9}$$

The above equality must be fulfilled outside supplier $\mathbf{X}$.

Suppose in (2.8) $\mathbf{T}$ is the potential vector, i.e. $\mathbf{T} = \nabla \Phi$. Then (2.8) is written as

$$\nabla \bullet \mathbf{T} = \Delta \Phi = \nabla \bullet \mathbf{X},$$

Then
$$\Phi = \Gamma * \nabla_D \bullet \mathbf{X} \tag{2.10}$$

Outside the supplier, like in (2.9), one can write down
$$\hat{\mathbf{U}} = \nabla \Phi$$

For the plane case, the solution (2.10) looks like

$$\Phi = \frac{1}{2\pi} \ln \sqrt{x^2 + y^2} * (\nabla \bullet \mathbf{X}) \tag{2.11}$$

According to (1.8), we obtain

$$\hat{U}_x = \frac{1}{2\pi}(\nabla \bullet \mathbf{X}) * \frac{x}{x^2 + y^2}$$
$$\hat{U}_y = \frac{1}{2\pi}(\nabla \bullet \mathbf{X}) * \frac{y}{x^2 + y^2} \tag{2.12}$$

So long as
$$\Delta \hat{U}_x = 0, \quad \Delta \hat{U}_y = 0,$$
then
$$C_0 \hat{\mathbf{U}} + C_1 \tilde{\mathbf{U}} \tag{2.13}$$

will be a solution of the equation (2.7) rather than that of the elastic equilibrium (2.4). Here
$$\tilde{\mathbf{U}} = \{\hat{U}_x, \ 0\} \tag{2.14}$$

$C_1$ and $C_2$ are the arbitrary constants. Divergence of the vector $\tilde{\mathbf{U}}$ is written as

$$\nabla \bullet \tilde{\mathbf{U}} = \frac{1}{2\pi}(\nabla \bullet \mathbf{X}) * \frac{y^2 - x^2}{(x^2 + y^2)^2} \tag{2.15}$$

Vector $\tilde{\mathbf{U}}$ is neither potential nor vortex, since $\nabla \bullet \tilde{\mathbf{U}} \not\equiv 0, \ \nabla \times \tilde{\mathbf{U}} \not\equiv 0$

Substituting $\tilde{\mathbf{U}}$ into (2.4) one can be convinced that $\tilde{\mathbf{U}}$ is its solution, but only at one value $a = 1$, which is the same as $\nu = \infty$. Such a value does not correspond to real materials. In order to obtain a solution for the arbitrary $a$, add to (1.23) the potential vector $\nabla \Psi$, such that its divergence is equal to (2.15)

$$\Delta \Psi = \frac{1}{2\pi}(\nabla \bullet \mathbf{X}) * \frac{y^2 - x^2}{(x^2 + y^2)^2} \tag{2.16}$$

Then the solution (2.4) is the vector
$$\mathbf{U} = \tilde{\mathbf{U}} - (1-a)\nabla \Psi \tag{2.17}$$

and the general solution (2.4) is written as ( $C_0$ and $C_1$ are arbitrary constants)

$$\mathbf{T} = C_0 \hat{\mathbf{U}} + C_1 \mathbf{U} \tag{2.18}$$

The method is demonstrated below by the solution of the problems of the two-dimensional theory of elasticity for dislocation, a crack in the infinite specimen and a surface crack. The above method has two advantages as compared to the method of the functions of the complex variable. Firstly, it is useful for three-dimensional problems and can be extended to nonstationary problems as well as those with a plastic displacement. Such problems cannot be solved in principle by means of the method of the functions of the complex variable. Secondly, the method makes is possible to calculate elastic energy analytically. The reader is referred to [2] for the solution of a three-dimensional problem obtained by the above method.

### 3. Dislocation

Let us consider dislocation as an insert of an additional atomic half plane along the positive part of the axis of ordinates. It is equivalent to the case of the displacement source of the X-type being located on the positive part of the axis of ordinates

$$X_x = b\,\Theta(x)\,\Theta(y)$$
$$X_y = 0$$

Here $b$ is the Burgers vector. The divergence X is

$$\nabla_D \bullet \mathbf{X} = b\,\delta(x)\Theta(y) = b\,\delta_{+y} \tag{3.1}$$

Hence, using (2.11) we derive

$$\Phi = \frac{b}{2\pi} \int_0^A \ln\sqrt{x^2 + (y-t)^2}\,dt$$

Here the upper limit is chosen as a finite value, since at $A \to \infty$ the integral diverges. Then by (2.12) we obtain at $A \to \infty$:

$$\hat{U}_x = \frac{b}{2\pi} \operatorname{arctg}\frac{y}{x}$$
$$\hat{U}_y = \frac{b}{2\pi} \ln\sqrt{x^2 + y^2}$$

The arbitrary constant vector, which does not contribute to deformation, can be added to $\hat{\mathbf{U}}$

$$\hat{\varepsilon}_{xx} = -\hat{\varepsilon}_{yy} = -\frac{b}{2\pi}\frac{y}{x^2+y^2}$$
$$\hat{\varepsilon}_{xy} = -\frac{b}{2\pi}\frac{x}{x^2+y^2}$$

Vector $\tilde{\mathbf{U}}$ is determined according to (2.14). Using (2.16), (3.1) we get

$$\Delta\Psi = -\frac{b}{2\pi}\frac{y}{x^2+y^2} \tag{3.2}$$

We shall seek for the solution of (3.2) using the method of separation of variables. In the polar coordinates $x = r\cos\varphi$, $y = r\sin\varphi$ (3.2) has the form

$$\frac{\partial^2 \Psi}{\partial r^2} + \frac{1}{r}\frac{\partial \Psi}{\partial r} + \frac{1}{r^2}\frac{\partial^2 \Psi}{\partial \varphi^2} = -\frac{b}{2\pi}\frac{\sin\varphi}{r} \tag{3.3}$$

Let us represent the unknown function in the form
$$\Psi(r,\varphi) = R(r)\sin\varphi$$
Then (3.3) will be written as
$$\frac{d^2 R}{dr^2} = -\frac{b}{2\pi}\frac{1}{r}$$
Integrating we get

$$R = -\frac{b}{2\pi}[r \ln r - r + Cr + C_1]$$

Having chosen the values of the arbitrary constants $C = 1$, $C_1 = 0$ we obtain a particular solution (3.3)

$$\Psi = -\frac{b}{2\pi} r \ln r \sin\varphi = -\frac{b}{2\pi} y \ln\sqrt{x^2 + y^2}$$

Hence

$$\nabla\Psi = -\frac{b}{4\pi}\left\{\frac{xy}{x^2+y^2},\ \ln\sqrt{x^2+y^2} + \frac{y^2}{x^2+y^2}\right\}$$

Vector **U** will be written as

$$U_x = \frac{b}{2\pi}\left[-\arctan\frac{y}{x} + \frac{1-a}{2}\frac{xy}{x^2+y^2}\right] = \frac{b}{2\pi}\left[-\varphi + \frac{1-a}{4}\sin 2\varphi\right]$$

$$U_y = \frac{b}{2\pi}\frac{1-a}{2}\left[\ln\sqrt{x^2+y^2} + \frac{y^2}{x^2+y^2}\right] = \frac{b}{2\pi}\frac{1-a}{2}[\ln r + \sin^2\varphi]$$

According to (2.16) vector **T** will be written as

$$T_x = \frac{b}{2\pi}\left[(C_0 - C_1)\varphi + C_1\frac{1-a}{4}\sin 2\varphi\right]$$

$$T_y = \frac{b}{2\pi}\left[\left(C_0 + \frac{1-a}{2}C_1\right)\ln r + C_1\frac{1-a}{2}\sin^2\varphi\right]$$

(3.4)

The arbitrary constants must be determined from supplementary conditions. For the above problem these conditions will be

1) In tracing the dislocation line along the boundary the displacement vector gains the increment $\mathbf{b} = b\{1,0\}$;
2) There must be no concentrated force on the dislocation line.

The increment of vector **T** in tracing the origin of the coordinates, where the dislocation is situated, is written as

$$\oint d\mathbf{T} = b\{C_0 - C_1, 0\} = b\{1,0\}$$

Hence

$$C_0 - C_1 = 1 \tag{3.5}$$

The stress tensor for (3.4) is written as

$$\sigma_{xx} = \frac{G}{\pi b}\frac{\sin\varphi}{r}\{-C_0 + C_1[2(a-1)\cos^2\varphi + (a-2)]\}$$

$$\sigma_{yy} = \frac{G}{\pi b}\frac{\sin\varphi}{r}\{C_0 + C_1[a - 2(a-1)\cos^2\varphi]\}$$

$$\sigma_{xy} = \frac{G}{\pi b}\frac{\cos\varphi}{r}\{C_0 + C_1[2(a-1)\cos^2\varphi + a]\}$$

The force acting on the arc segment $rd\varphi$ is written in the Cartesian system of coordinates as

$$d\mathbf{F} = \{\sigma_{xx}n_x + \sigma_{xy}n_y,\ \sigma_{xy}n_x + \sigma_{yy}n_y\}rd\varphi$$

Here $\mathbf{n} = \{\cos\varphi,\ \sin\varphi\}$ is normal to the segment $rd\varphi$. Having integrated $d\mathbf{F}$ with respect to $\varphi$ from 0 to $2\pi$ we obtain the concentrated force

$$\mathbf{F} = \int_0^{2\pi} d\mathbf{F} = 2Gb\{0,\ C_0 + C_1\}$$

In equilibrium it is zero, hence

$C_0 + C_1 = 0$

From the above and (3.5) we get

$C_0 = -C_1 = 1/2$.

From (3.4) we obtain

$$T_x = \frac{b}{2\pi}\left[\varphi + \frac{1-a}{4}\sin 2\varphi\right]$$

$$T_y = \frac{b}{2\pi}\left[-a\ln r + \frac{1-a}{2}\sin^2\varphi\right],$$

which is a well-known expression for the elastic field of the edge dislocation [4,5].

Application of the principle of the minimum elastic energy to determine the arbitrary constants $C_0$ and $C_1$ gives the same result. Really, the minimum requires the derivatives of the total energy with respect to the generalized coordinates to be equal to zero.

**4. A Flat Crack**

The problem on the flat crack has been solved before by the method of the functions of the complex variable [4,5,6]. Here we shall find a solution of the same problem using generalized functions. Let us divide the elastic field in the specimen into an external field and a peculiar field of the crack. The external field is the field in the loaded specimen that has not been cut. The external load is the uniaxial tension along the axis of ordinates due to the external tension $\sigma^{ex}$. It produces deformation

$$\varepsilon_{xx}^{ex} = -\frac{\sigma^{ex}(1-2a)}{2G(3-4a)}, \qquad \varepsilon_{yy}^{ex} = \frac{\sigma^{ex}(1-a)}{G(3-4a)}, \qquad \varepsilon_{xy}^{ex} = 0 \qquad (4.1)$$

If one makes a cut in a loaded specimen leaving the edges free, then the crack will open and an additional deformation $\varepsilon$ to (4.1) will appear. Let us call it a peculiar field of the crack or just a crack field. The total field of the deformation $\varepsilon^f$ is

$\varepsilon^f = \varepsilon^{ex} + \varepsilon$

The cut is located on the abscissa axis on the segment $(-L, L)$, and we shall designate it by $M$, and the arbitrary point on the plane $xy$ by $P$. The boundary conditions are

$\sigma^{ex} + \sigma_{yy} = 0, \quad \sigma_{xy} = 0 \quad$ at $P \in M$

(4.2)

$\sigma \to 0$ when $P \to \infty$

On $M$ normal to the segment displacement changes by a jump

$[T_y]_M = F(x)$ when $P \in M$

therefore, the displacement source $M$ is written as

$X = \frac{1}{2}F(x)\text{sign } y \quad$ when $P \in M$

Component $T_x$ is continuous. Divergence in the generalized sense is written as

$$\nabla_D \bullet \mathbf{T} = \frac{\partial}{\partial x}T_x + D_y T_y = \frac{\partial}{\partial x}T_x + \frac{\partial}{\partial y}T_y + F(x)\delta_M = \nabla \bullet \mathbf{T} + F(x)\delta_M$$

Therefore, the divergence source is

$\nabla \bullet X_0 = F(x)\delta_M$

The unknown function $F(x)$ sets opening of the crack.

Since we are dealing with a crack of a normal opening, we shall have to do with the equation (2.7), which is written as

$\Delta(\nabla \bullet \mathbf{T}) = \Delta(F(x)\delta_M)$.

According to (2.8), for the first component in the right side we have

$$\nabla \bullet \mathbf{T} = F(x)\delta_M$$

According to (2.11), we obtain

$$\Phi = \frac{1}{2\pi}\ln\sqrt{x^2+y^2} * F(x)\delta_M = \frac{1}{2\pi}\int_{-L}^{L}F(t)\ln\sqrt{(x-t)^2+y^2}\,dt \qquad (4.3)$$

Hence, taking into account (1.3) and (2.12) we get

$$\hat{U}_x = \frac{1}{2\pi}F(x)\delta_M * \frac{x}{x^2+y^2} = \frac{1}{2\pi}\int_{-L}^{L}f(t)\ln\sqrt{(x-t)^2+y^2}\,dt$$

$$\hat{U}_y = \frac{1}{2\pi}F(x)\delta_M * \frac{y}{x^2+y^2} = \frac{1}{2\pi}\int_{-L}^{L}f(t)\arctan\frac{x-t}{y}\,dt \qquad (4.4)$$

Here

$$f(x) = -\frac{dF(x)}{dx} \qquad (4.5)$$

At $y \to 0$

$$\arctan\frac{x-t}{y} \to \frac{\pi}{2}\mathrm{sign}(x-t)\mathrm{sign}(y)$$

therefore, according to (1.24), the displacement $\hat{U}_y$ on segment $M$ looks like

$$\hat{U}_y(x,0) = \lim_{y\to 0}\frac{1}{2\pi}\int_{-L}^{L}f(t)\arctan\frac{x-t}{y}\,dt =$$

$$= \frac{1}{2}\mathrm{sign}(y)\int_{-L}^{L}f(t)\mathrm{sign}(x-t)\,dt = -\frac{1}{2}F(x)\mathrm{sign}(y) \qquad (4.6)$$

Therefore

$$\Delta\hat{U}_y = -\frac{1}{2}F(x)D_y\,\delta(y) \qquad (4.7)$$

Differentiating (4.6) we obtain deformation on $M$:

$$\hat{\varepsilon}_{yy}(x,0) = -\hat{\varepsilon}_{xx}(x,0) = \frac{\partial \hat{U}_y(x,0)}{\partial y} = -\frac{1}{2}F(x)\delta_M = -\frac{1}{2\pi}\int_{-L}^{L}\frac{f(t)}{x-t}\,dt \qquad (4.8)$$

The diverging integrals are understood in the sense of the principal value, namely

$$\int_{-L}^{L}\frac{f(t)}{x-t}\,dt = \lim_{\rho \to 0}\left[\int_{-L}^{x-\rho}\frac{f(t)}{x-t}\,dt + \int_{x+\rho}^{L}\frac{f(t)}{x-t}\,dt\right] \qquad (4.9)$$

According to (2.16),

$$\Delta\Psi = \frac{1}{2\pi}F(x)\delta_M * \frac{y^2-x^2}{(x^2+y^2)^2} = \frac{1}{2\pi}F(x)\delta_M * \frac{\partial}{\partial x}\frac{x}{x^2+y^2} =$$

$$= \frac{1}{2\pi}\frac{\partial}{\partial x}F(x)\delta_M * \frac{x}{x^2+y^2} = \frac{1}{2\pi}\int_{-L}^{L}f(t)\frac{x-t}{(x-t)^2+y^2}\,dt = \hat{\varepsilon}_{xx} \qquad (4.10)$$

The solution (4.10) is derived by separating the variables in the same way as for (3.3):

$$\Psi = -\frac{1}{4\pi}\int_{-L}^{L}f(t)(x-t)\ln\sqrt{(x-t)^2+y^2}\,dt$$

Gradient $\Psi$ is written as

$$\nabla\Psi = -\frac{1}{4\pi}\left\{\int_{-L}^{L}f(t)\ln\sqrt{(x-t)^2+y^2} + \frac{(x-t)^2}{(x-t)^2+y^2}\,dt\,,\int_{-L}^{L}f(t)\frac{(x-t)y}{(x-t)^2+y^2}\,dt\right\} \qquad (4.11)$$

Vector $\tilde{\mathbf{U}}$ is obtained from (2.14). Vectors $\tilde{\mathbf{U}}$ and $\nabla\Psi$ are continuous when passing through segment $M$ in contrast to $\hat{\mathbf{U}}$ and do not contribute to the value of the crack opening. Only component $\hat{U}_y$ of vector $\hat{\mathbf{U}}$ has a rupture on segment $M$. Outside segment $M$ each component $\hat{\mathbf{U}}$ satisfies the Laplace equation, so one can write

$$\Delta \hat{U}_y = F(x)\delta_M$$

Then a common solution is given by (2.18). The deformations produced by vector $\hat{\mathbf{U}}$, according to (2.3), are as follows

$$\hat{\varepsilon}_{xx} = -\hat{\varepsilon}_{yy} = \frac{1}{2\pi} f(x)\delta_M * \frac{x}{x^2 + y^2} = \frac{1}{2\pi}\int_{-L}^{L} f(t)\frac{x-t}{(x-t)^2 + y^2}\,dt$$

$$\hat{\varepsilon}_{xy} = \frac{1}{2\pi} f(x)\delta_M * \frac{y}{x^2 + y^2} = \frac{1}{2\pi}\int_{-L}^{L} f(t)\frac{y}{(x-t)^2 + y^2}\,dt$$

(4.12)

those produced by vector $\tilde{\mathbf{U}}$:

$$\tilde{\varepsilon}_{xx} = \hat{\varepsilon}_{xx}, \quad \tilde{\varepsilon}_{yy} = 0, \quad \tilde{\varepsilon}_{xy} = \frac{1}{2}\hat{\varepsilon}_{xy}$$

and those caused by vector $\nabla\Psi$:

$$\varepsilon_{xx}^{\Psi} = \frac{1}{2}\hat{\varepsilon}_{xx} + \omega$$

$$\varepsilon_{yy}^{\Psi} = \frac{1}{2}\hat{\varepsilon}_{xx} - \omega$$

$$\varepsilon_{xy}^{\Psi} = -\frac{\hat{\varepsilon}_{xy}}{2} + \omega_1$$

Here

$$\omega = \frac{1}{2\pi}\int_{-L}^{L} f(t)\frac{(x-t)y^2}{[(x-t)^2 + y^2]^2}\,dt = \frac{y}{2}\frac{\partial(\hat{\varepsilon}_{xx})}{\partial y}$$

$$\omega_1 = \frac{1}{2\pi}\int_{-L}^{L} f(t)\frac{y^3}{[(x-t)^2 + y^2]^2}\,dt$$

(4.13)

The deformation tensor produced by the displacement vector $\mathbf{T}$ by (2.17), (2.18), (4.4) and (4.11) is written as

$$\varepsilon_{xx} = C_0\hat{\varepsilon}_{xx} + C_1[\hat{\varepsilon}_{xx} - (1-a)\varepsilon_{xx}^{\Psi}]$$

$$\varepsilon_{yy} = C_0\hat{\varepsilon}_{yy} - C_1(1-a)\varepsilon_{yy}^{\Psi}$$

$$\varepsilon_{xy} = C_0\hat{\varepsilon}_{xy} + C_1\left[\frac{1}{2}\hat{\varepsilon}_{xy} - (1-a)\varepsilon_{xy}^{\Psi}\right]$$

$$\varepsilon_{xx} + \varepsilon_{yy} = C_1 a\hat{\varepsilon}_{xx}$$

(4.14)

At $P \in M$ we have

$$\omega(M) = 0$$

$$\varepsilon_{xx}^{\Psi}(M) = \varepsilon_{yy}^{\Psi}(M) = \frac{1}{2}\hat{\varepsilon}_{xx}(M)$$

(4.15)

Differentiating (4.6) with respect to $x$ we obtain

$$\hat{\varepsilon}_{xy}(M) = \frac{1}{2} f(x)\operatorname{sign} y$$

Hence, the shear stress on the surface of the crack from (4.2) and (4.14) is written as

$$\sigma_{xy} = G\left(C_0 + \frac{C_1}{2}\right)f(x)\operatorname{sign} y \equiv 0$$

It is possible only at
$$C_1 = -2C_0 \tag{4.16}$$

The stress tensor of the peculiar field of the crack is written as
$$\sigma_{xx} = 2(1-a)GC_0\left[-\hat{\varepsilon}_{xx} + 2\omega\right]$$
$$\sigma_{yy} = 2(1-a)GC_0\left[\hat{\varepsilon}_{yy} - 2\omega\right] \tag{4.17}$$
$$\sigma_{xy} = 4GC_0(1-a)\varepsilon_{xy}^{\Psi}$$

Normal stresses on the surface of the crack taking into account (4.15) will be
$$\sigma_{yy}(M) = 2(1-a)GC_0\left[\hat{\varepsilon}_{yy}(M) - 2\omega(M)\right] = -(1-a)C_0 \frac{G}{\pi}\int_{-L}^{L} \frac{f(t)}{x-t}dt = -\sigma^{ex}$$

Hence
$$\int_{-L}^{L} \frac{f(t)}{x-t}dt = \frac{\pi\sigma^{ex}}{C_0 G(1-a)} \tag{4.18}$$

The solution of this singular equation is shown in [5], where it is demonstrated that there are three solutions (4.18). They differ in the behaviour near the ends of segment $M$. The first solution is limited at both ends. It requires a condition which cannot be satisfied in this case
$$\int_{-L}^{L} \frac{t\sqrt{L^2 - t^2}}{(x-t)}dt = 0$$

It means that such solution does not exist. The second solution is limited at one end of segment $M$ and unlimited at the other. The above solution is related to the wedge-shaped crack.

The third solution is not limited at either ends of segment $M$. It looks like
$$f(x) = -\frac{1}{\sqrt{L^2 - x^2}}\left[\frac{1}{\pi}\int_{-L}^{L}\frac{\sqrt{L^2 - t^2}}{t - x}dt + C\right]$$

Here $C$ is the arbitrary constant. Integrating we obtain
$$f(x) = -\frac{x - C}{\sqrt{L^2 - x^2}}$$

Its primitive by (4.5) is
$$F(x) = -\sqrt{L^2 - x^2} + C\arcsin\frac{x}{L}$$

The solution must be symmetrical with respect to the axis of ordinates. For this it is necessary that $C = 0$. Finally, we get
$$F(x) = -\sqrt{L^2 - x^2}$$
$$f(x) = -\frac{x}{\sqrt{L^2 - x^2}} \tag{4.19}$$

Then
$$\int_{-L}^{L}\frac{f(t)}{x-t}dt = -\pi \tag{4.20}$$

From (4.8) we obtain
$$\hat{\varepsilon}_{yy}(x,0) = -\hat{\varepsilon}_{xx}(x,0) = \frac{1}{2}$$

From (4.16), (4.18) and (4.20) we get

$$C_0 = -\frac{\sigma^{ex}}{(1-a)G} \tag{4.21}$$

$$C_1 = -2C_0$$

The value of the crack opening will be

$$T_y = \frac{\sigma^{ex}}{(1-a)G}\sqrt{L^2 - x^2}\,\text{sign}(y)$$

The deformation tensor will be written as

$$\varepsilon_{xx} = \frac{\sigma^{ex}}{(1-a)G}\left[a\hat{\varepsilon}_{xx} - 2(1-a)\omega\right]$$

$$\varepsilon_{yy} = -\frac{\sigma^{ex}}{(1-a)G}\left[a\hat{\varepsilon}_{xx} + 2(1-a)\omega\right] \tag{4.22}$$

$$\varepsilon_{xy} = -\frac{2\sigma^{ex}}{G}\varepsilon_{xy}^{\Psi}$$

$$\varepsilon_{xx} + \varepsilon_{yy} = \frac{2a\sigma^{ex}}{(1-a)G}\hat{\varepsilon}_{xx}$$

The stress tensor will be written as

$$\sigma_{xx} = 2\sigma^{ex}\left[\hat{\varepsilon}_{xx} - 2\omega\right]$$

$$\sigma_{yy} = 2\sigma^{ex}\left[\hat{\varepsilon}_{xx} + 2\omega\right] \tag{4.23}$$

$$\sigma_{xy} = -4\sigma^{ex}\varepsilon_{xy}^{\Psi}$$

Since $|F(x)| \leq L$, then, according to (4.4), at $r = \sqrt{x^2 + y^2} \to \infty$ the displacement and deformations are as follows

$$\hat{\mathbf{U}} \sim \frac{1}{r}, \quad \hat{\varepsilon} \sim \frac{1}{r^2} \to 0 \tag{4.24}$$

## 5. A Crack on the Surface

In the problem on the crack examined above let us add one more cut along the axis of ordinates. Due to the stresses of the peculiar field of the crack a displacement will occur in it and an additional elastic field $\mathbf{U}^c$ will appear. Let us call the latter a compensation field. Its deformation is $\varepsilon^c$ and its stress is $\sigma^c$. The following boundary conditions must be satisfied on the axis of ordinates

$$\sigma_{xx}(0,y) + \sigma_{xx}^c(0,y) = 0, \quad \sigma_{xy}(0,y) + \sigma_{xy}^c(0,y) = 0$$

and on segment $M$

$$\sigma_{yy}(x,0) + \sigma^{ex} + \sigma_{yy}^c(x,0) = 0$$

$$\sigma_{xy}(x,0) + \sigma_{xy}^c(x,0) = 0$$

At infinity $\sigma^c \to 0$. On the axis of ordinates the total stresses are equal to zero, one half can be removed and we get a problem on a half plane with a surface crack.

At $x = 0$ in (4.17) $\sigma_{xy} = 0$, since under the integral $\varepsilon_{xy}^{\Psi}$ in (4.13) there is an odd function in $t$. For $\hat{\varepsilon}_{xx}$ we have

$$\hat{\varepsilon}_{xx}(0,y) = -\frac{1}{2\pi}\int_{-L}^{L}\frac{t^2}{\sqrt{L^2-t^2}}\frac{dt}{t^2+y^2} = \frac{1}{2}\left(\frac{|y|}{\sqrt{L^2+y^2}} - 1\right) \tag{5.1}$$

Integration in (5.1) is performed through passing to complex variables

$$\hat{\varepsilon}_{xx} + i\hat{\varepsilon}_{xy} = \frac{1}{2\pi}\left\{-\int_{-L}^{L}\frac{dt}{\sqrt{L^2-t^2}} + \bar{z}\int_{-L}^{L}\frac{dt}{\sqrt{L^2-t^2}(\bar{z}-t)}\right\} = \frac{i\bar{z}}{2\sqrt{L^2-\bar{z}^2}} - \frac{1}{2} \quad (5.2)$$

Here $\bar{z} = x - i y$. Assuming in (5.2) that $x = 0$ we obtain (5.1). The integrand in (5.1) is even in $y$. To retain evenness for the integration result in (5.2), it is necessary to introduce the function sign $y$. Besides, if the function sign $y$ is not introduced, then the function

$$\frac{y}{\sqrt{L^2+y^2}} - 1 \to -2, \quad y \to -\infty$$

which contradicts (4.24). Introduction of sign $y$ is due to the fact that the square root is a two valued function, and its value should be chosen based on the physical sense. There is an equality $y \, \text{sign}\, y = |y|$.

According to (4.13) and (5.1),

$$\omega(0, y) = \left(\frac{y}{2}\frac{d\hat{\varepsilon}_{xx}}{dy}\right)\bigg|_{x=0} = -\frac{L^2|y|}{4(L^2+y^2)^{3/2}}$$

The stress on the axis of ordinates caused by the crack field on segment $M$ by (4.23) is

$$\sigma_{xx}((0,y)) = 2\sigma^{ex}[-\hat{\varepsilon}_{xx}(0,y) + 2\omega(0,y)] = \sigma^{ex}\left[\frac{|y|}{\sqrt{L^2+y^2}} - 1 - \frac{L^2|y|}{(L^2+y^2)^{3/2}}\right] =$$

$$= -\frac{\sigma^{ex}}{\pi}\int_{-L}^{L}\frac{t^2}{\sqrt{L^2-t^2}}\frac{t^2-y^2}{(t^2+y^2)^2}dt \quad (5.3)$$

The stress (5.3) is represented in two forms: explicit and integral. Let us introduce a source of displacement into the axis of ordinates

$$\chi = 2\pi P(y)\delta_y \quad (5.4)$$

As in the previous section, $P(y)$ is the unknown function. It should be chosen in such a way that the stress in (5.3) is compensated. The source of displacement (5.4) corresponds to the field of displacement

$$\mathbf{T}^c = C_0 \hat{\mathbf{U}}^c + C_1 \mathbf{U}^c$$

Here

$$\hat{U}_x^c = \frac{1}{2\pi}\int_{-\infty}^{\infty} p(t)\arctan\frac{y-t}{x}dt$$

$$\hat{U}_y^c = \frac{1}{2\pi}\int_{-\infty}^{\infty} p(t)\ln\sqrt{x^2+(y-t)^2}\,dt$$

Vector $\mathbf{U}^c$ is written as

$$\mathbf{U}^c = \tilde{\mathbf{U}}^c - (1-a)\nabla \Psi^c$$

where

$$\tilde{\mathbf{U}} = \{0, \; \hat{U}_y^c\}$$

$$\Psi^c = \frac{1}{4\pi}\int_{-\infty}^{\infty} P(t)(y-t)\ln\sqrt{x^2+(y-t)^2}\,dt$$

$$p(y) = -\frac{\partial P(y)}{\partial y} \quad (5.5)$$

The deformation is written in the form :

$$\hat{\varepsilon}_{yy}^c = -\hat{\varepsilon}_{xx}^c = \frac{1}{2\pi} \int_{-\infty}^{\infty} p(t) \frac{y-t}{x^2 + (y-t)^2} dt$$

$$\hat{\varepsilon}_{xy}^c = \frac{1}{2\pi} \int_{-\infty}^{\infty} p(t) \frac{x}{x^2 + (y-t)^2} dt \tag{5.6}$$

$$\varepsilon_{xx}^{\Psi c} = \frac{1}{2} \hat{\varepsilon}_{yy}^c - \omega^c$$

$$\varepsilon_{yy}^{\Psi c} = \frac{1}{2} \hat{\varepsilon}_{yy}^c + \omega^c$$

$$\varepsilon_{xy}^{\Psi c} = -\frac{1}{2} \hat{\varepsilon}_{xy}^c + \omega_1^c \tag{5.7}$$

$$\omega^c = \frac{1}{2\pi} \int_{-\infty}^{\infty} p(t) \frac{(y-t)x^2}{\left[x^2 + (y-t)^2\right]^2} dt$$

$$\omega_1^c = \frac{1}{2\pi} \int_{-\infty}^{\infty} p(t) \frac{x^3}{\left[x^2 + (y-t)^2\right]^2} dt$$

Let us assume the unknown constants $C_0$, $C_1$ to be equal (4.21). It does not reduce the generality of the problem, so we have the unknown function $P(y)$ at our disposal. The stresses will be as follows

$$\sigma_{xx}^c = 2\sigma^{ex} \left[\hat{\varepsilon}_{yy}^c + 2\omega^c\right]$$

$$\sigma_{yy}^c = 2\sigma^{ex} \left[\hat{\varepsilon}_{yy}^c - 2\omega^c\right] \tag{5.8}$$

$$\sigma_{xy} = 4\sigma^{ex} \varepsilon_{xy}^{\Psi}$$

From (5.5) and (5.7) it follows that

$$\varepsilon_{xy}^{\Psi c}(0, y) = 0$$

$$\omega^c(0, y) = 0$$

Then according to (5.8), we obtain

$$\sigma_{xy}^c(0, y) = 0$$

$$\sigma_{xx}^c(0, y) = 2\sigma^{ex} \hat{\varepsilon}_{yy}^c(0, y)$$

From (5.1) and (5.3) it follows

$$\sigma_{xx}(0, y) + \sigma_{xx}^c(0, y) = 0$$

Hence, taking into account (4.23), (5.3) and (5.6)

$$\int_{-L}^{L} \frac{t^2}{\sqrt{L^2 - t^2}} \frac{t^2 - y^2}{(t^2 + y^2)^2} dt = \int_{-\infty}^{\infty} \frac{p(t)}{t - y} dt$$

We have derived a singular integral equation for the unknown function $p(y)$. Its solution is known [5]:

$$p(y) = -\frac{1}{\pi^2} \int_{-\infty}^{\infty} \int_{-L}^{L} \frac{t^2}{\sqrt{L^2 - t^2}} \frac{t^2 - t_1^2}{(t^2 + t_1^2)^2} dt \frac{dt_1}{t_1 - y}$$

Let us take the integral

$$\int_{-\infty}^{\infty} \frac{t^2 - t_1^2}{(t^2 + t_1^2)^2} \frac{dt_1}{t_1 - y} = -2\pi \frac{ty}{t^2 + y^2} \operatorname{sign} t$$

Then

$$p(y) = \frac{2y}{\pi} \int_{-L}^{L} \frac{t^3}{\sqrt{L^2 - t^2}} \frac{\operatorname{sign} t \, dt}{t^2 + y^2} \tag{5.9}$$

$$\int_{-L}^{L} \frac{t^3}{\sqrt{L^2 - t^2}} \frac{dt}{t^2 + y^2} = \frac{2}{\sqrt{L^2 + y^2}} \operatorname{arctanh} \frac{L}{\sqrt{L^2 + y^2}}$$

As a result, we get

$$p(y) = -\frac{4y}{\pi\sqrt{L^2 + y^2}} \operatorname{arctanh} \frac{L}{\sqrt{L^2 + y^2}}$$

A jump of displacement on the additional cut is written, according to (5.5), as

$$P(y) = -\frac{2}{\pi} \left[ \sqrt{L^2 + y^2} \ln \frac{\sqrt{L^2 + y^2} + L}{\sqrt{L^2 + y^2} - L} + 2L \ln \frac{y}{L} \right]$$

The arbitrary integration constant is dropped. The displacement on the additional cut has a logarithmic divergence at infinity.

The deformations of the compensation field $\mathbf{T}^c$ on the abscissa axis are

$$\hat{\varepsilon}_{xy}^c = \frac{1}{2\pi} \int_{-\infty}^{\infty} p(t) \frac{x}{x^2 + t^2} dt = 0$$

due to $p(y)$ being odd. The stress $\sigma_{yy}^c$ on the abscissa is written as

$$\sigma_{yy}^c(x,0) = 4\frac{1-2a}{1-a} \sigma^{ex} \hat{\varepsilon}_{yy}^c(x,0)$$

The integral representation $p(y)$ is used for

$$\hat{\varepsilon}_{yy}^c = -\frac{1}{2\pi} \int_{-\infty}^{\infty} \frac{t\, p(t)}{x^2 + t^2} dt = -\frac{1}{\pi^2} \int_{-L}^{L} \int_{-\infty}^{\infty} \frac{t^3 \operatorname{sign} t}{\sqrt{L^2 - t^2}} \frac{t_1}{x^2 + t_1^2} \frac{dt_1}{t^2 + t_1^2} dt = 0$$

since under the integral there is an odd function in $t_1$. The compensation field does not produce stress on the crack surface.

Thus, we have found an accurate solution of the problem on the flat crack emerging on the free surface of a semi-infinite body.

### 6. Elastic Energy of the Crack in the Infinite Specimen

Deformation of an infinite specimen with a crack is written as

$$\bar{\varepsilon} = \varepsilon + \varepsilon^{ex} \tag{6.1}$$

The density of the elastic energy of a specimen with a crack is [8]:

$$\bar{w} = \frac{1-2a}{2a} G(\bar{\varepsilon}_{xx} + \bar{\varepsilon}_{yy})^2 + G[\bar{\varepsilon}_{xx}^2 + 2\bar{\varepsilon}_{xy}^2 + \bar{\varepsilon}_{yy}^2] = w^{ex} + w + w^i \tag{6.2}$$

The bar above the symbol denotes that the value is related to the total field of elastic deformations. Here

$$w^{ex} = \frac{1-2a}{2a} G(\varepsilon_{xx}^{ex} + \varepsilon_{yy}^{ex})^2 + G[(\varepsilon_{xx}^{ex})^2 + 2(\varepsilon_{xy}^{ex})^2 + (\varepsilon_{yy}^{ex})^2] \tag{6.3}$$

is the density of the energy of external stresses

$$w^i = \frac{1-2a}{a} G(\varepsilon_{xx}^{ex} + \varepsilon_{yy}^{ex})(\varepsilon_{xx} + \varepsilon_{yy}) + 2G[\varepsilon_{xx}^{ex}\varepsilon_{xx} + 2\varepsilon_{xy}^{ex}\varepsilon_{xy} + \varepsilon_{yy}^{ex}\varepsilon_{yy}] \tag{6.4}$$

is the density of the energy of the interaction between the external field and the crack field

$$w = \frac{1-2a}{2a} G(\varepsilon_{xx} + \varepsilon_{yy})^2 + G[(\varepsilon_{xx})^2 + (\varepsilon_{yy})^2 + 2(\varepsilon_{xy})^2] =$$
$$= 2a(1-2a)C_0^2 G\hat{\varepsilon}_{xx}^2 + G\left[|\nabla T_x|^2 + |\nabla T_y|^2\right] \tag{6.5}$$

is the density of the energy of the crack field. Having integrated the energy density over the whole specimen we shall obtain the total elastic energy of the specimen with a crack

$$\overline{W} = \int_V \overline{w} \, dV \tag{6.6}$$

It consists of the elastic energy of the external field

$$W^{ex} = \int_V w^{ex} \, dV \tag{6.7}$$

the elastic energy of the interaction between the external field and the crack field

$$W^i = \int_V w^i \, dV \tag{6.8}$$

and the elastic energy of the crack

$$W = \int_V w \, dV \tag{6.9}$$

The external field energy (6.7) does not depend on the length of the crack, and during differentiation with respect to $L$ it reduces to zero. Besides, it is not limited in an infinite specimen, so to avoid its divergence, it should be subtracted from the total energy (6.6).

The surface area of the crack is $4L$. The surface energy of the crack $4\gamma L$ must be added to (6.6), here $\gamma$ is the surface energy density, and we shall obtain the potential energy of the specimen with a crack

$$W^P = \overline{W} + 4\gamma L \tag{6.10}$$

Having differentiated $W^P$ with respect to the crack length $2L$ we shall get a generalized force acting upon the crack in the form

$$g(L) = \frac{\partial W^P}{\partial (2L)} = \frac{1}{2} \frac{\partial \overline{W}}{\partial L} + 2\gamma \tag{6.11}$$

If $g(L) > 0$, then the crack tends to close, since with increasing length of the crack the energy (6.10) increases. If $g(L) < 0$, then the crack tends to grow, since with increasing length of the crack the energy (6.10) decreases. The condition $g(L) = 0$ is critical.

According to (2.3), (4.4) and the Green theorem

$$\int_V \left( \hat{\varepsilon}_{xy}^2 + \hat{\varepsilon}_{yy}^2 \right) dV = \int_V \left| \nabla \hat{U}_y \right|^2 dV = \int_S \hat{U}_y \frac{\partial \hat{U}_y}{\partial n} dS - \int_V \hat{U}_y \Delta \hat{U}_y \, dV =$$

$$= -\int_V \hat{U}_y D_y [F * \delta_M] dV = \int_{-L}^{L} F(x) \frac{\partial \hat{U}_y(x,0)}{\partial y} dx = \tag{6.12}$$

$$= \int_{-L}^{L} F(x) \hat{\varepsilon}_{yy}(x,0) dx = \frac{\pi}{4} L^2$$

Similarly

$$\int_V \left( \hat{\varepsilon}_{xy}^2 - \hat{\varepsilon}_{yy}^2 \right) dV = \int_V \left( \hat{\varepsilon}_{xy} - \hat{\varepsilon}_{yy} \right)\left( \hat{\varepsilon}_{xy} + \hat{\varepsilon}_{yy} \right) dV = -\int_V \left( \hat{\varepsilon}_{xy} + \hat{\varepsilon}_{xx} \right)\left( \hat{\varepsilon}_{xy} + \hat{\varepsilon}_{yy} \right) dV =$$

$$= -\int_V \nabla \hat{U}_x \nabla \hat{U}_y \, dV = -\int_S \hat{U}_y \frac{\partial \hat{U}_x}{\partial n} dS + \int_V \hat{U}_y \Delta \hat{U}_x \, dV = 0 \tag{6.13}$$

because $\Delta \hat{U}_x = 0$ across the whole plane $xy$. From (4.12), (6.12), (6.13) it follows that

$$\int_V \hat{\varepsilon}_{xx}^2 \, dV = \int_V \hat{\varepsilon}_{yy}^2 \, dV = \int_V \hat{\varepsilon}_{xy}^2 \, dV = \frac{\pi}{8} L^2 \tag{6.14}$$

By (2.18)
$$\mathbf{T} = C_0 \hat{\mathbf{U}} + C_1 \mathbf{U} = C_0 \hat{\mathbf{U}} - C_1(1-a)\nabla \Psi + C_1 \tilde{\mathbf{U}}$$
According to the Green theorem,

$$J_1 = \int_V |\nabla T_x|^2 + |\nabla T_y|^2 \, dV = -\int_V (T_x \Delta T_x + T_y \Delta T_y) dV + \int_S \left( T_x \frac{\partial T_x}{\partial n} + T_y \frac{\partial T_y}{\partial n} \right) dS =$$

$$= -\int_V (T_x \Delta T_x + T_y \Delta T_y) dV + \frac{1}{2}\int_S \frac{\partial (T_x^2 + T_y^2)}{\partial n} dS = -\int_V (T_x \Delta T_x + T_y \Delta T_y) dV + \frac{1}{2}\int_S \frac{\partial |\mathbf{T}|^2}{\partial n} dS \quad (6.15)$$

By (4.4) and (4.24) at larger $r = \sqrt{x^2 + y^2}$

$$|\mathbf{T}| \sim \frac{1}{r},$$

then

$$|\mathbf{T}|^2 \sim \frac{1}{r^2}, \text{ a } \frac{\partial |\mathbf{T}|^2}{\partial n} \sim \frac{1}{r^3}$$

Therefore, when the contour diameter $S \to \infty$

$$\int_S \frac{\partial |\mathbf{T}|^2}{\partial n} dS \to 0$$

There is another way of evaluating the integral. Due to symmetry with respect to the axes of the coordinates the function $|\mathbf{T}|^2$ will be an even function in both variables. Its derivative $\partial |\mathbf{T}|^2/\partial x$ - is odd in $x$ and even in $y$, $\partial |\mathbf{T}|^2/\partial y$ is even in $x$ and odd in $y$. Let us take as the integration contour a square with the center at the origin of the coordinates and the vertices at the points $(A, A)$, $(-A, A)$, $(A, -A)$, $(-A, -A)$. Then we have

$$\oint_S \frac{\partial |\mathbf{T}|^2}{\partial n} dS = \int_{-A}^{A} \frac{\partial |\mathbf{T}|^2}{\partial x}(A, y) dy + \int_{A}^{-A} \frac{\partial |\mathbf{T}|^2}{\partial y}(x, A) dx - \int_{A}^{-A} \frac{\partial |\mathbf{T}|^2}{\partial x}(-A, y) dy - \int_{-A}^{A} \frac{\partial |\mathbf{T}|^2}{\partial y}(x, -A) dx =$$

$$= \int_{-A}^{A} \frac{\partial |\mathbf{T}|^2}{\partial x}(A, y) dy + \int_{-A}^{A} \frac{\partial |\mathbf{T}|^2}{\partial x}(-A, y) dy - \int_{-A}^{A} \frac{\partial |\mathbf{T}|^2}{\partial y}(x, A) dx - \int_{-A}^{A} \frac{\partial |\mathbf{T}|^2}{\partial y}(x, -A) dx = 0$$

The above way of integration is the generalization of the notion of the principal value of the integral (4.9) applied for the one-dimensional case to the multidimensional case. The integral diverging along an arbitrary contour can converge along the symmetric path. We obtain from (6.15):

$$J_1 = \int_V |\nabla T_x|^2 + |\nabla T_y|^2 \, dV = -\int_V [T_x \Delta T_x + T_y \Delta T_y] dV \quad (6.16)$$

By (2.17), (2.18), (4.4) and (4.11) we have

$$T_x = C_0 \left[ -\hat{U}_x + 2(1-a)\frac{\partial \Psi}{\partial x} \right]$$

$$T_y = C_0 \left[ \hat{U}_y + 2(1-a)\frac{\partial \Psi}{\partial y} \right] \quad (6.17)$$

Hence (4.10):

$$\Delta T_x = 2C_0(1-a)\frac{\partial \hat{\varepsilon}_{xx}}{\partial x}$$

$$\Delta T_y = 2C_0(1-a)\frac{\partial \hat{\varepsilon}_{xx}}{\partial y} \quad (6.18)$$

We have the following equalities

$$\int_{-\infty}^{\infty} \hat{U}_x \frac{\partial \hat{\varepsilon}_{xx}}{\partial x} dx = \hat{U}_x \hat{\varepsilon}_{xx} \Big|_{-\infty}^{\infty} - \int_{-\infty}^{\infty} \hat{\varepsilon}_{xx}^2 dx = -\int_{-\infty}^{\infty} \hat{\varepsilon}_{xx}^2 dx$$

$$-\left[T_x \Delta T_x + T_y \Delta T_y\right] = 2C_0^2(1-a)\left\{\nabla[\Phi + 2(1-a)\Psi]\nabla(\Delta\Psi) - 2\hat{U}_x \frac{\partial \hat{\varepsilon}_{xx}}{\partial x}\right\}$$

From the above equation and from (6.16), (4.10) we obtain

$$\begin{aligned}J_1 &= -2C_0^2(1-a)G\iint_V\left\{\nabla[\Phi + 2(1-a)\Psi]\nabla(\Delta\Psi) - 2\hat{U}_x \frac{\partial \hat{\varepsilon}_{xx}}{\partial x}\right\}dV = \\ &= -2C_0^2(1-a)G\iint_V\left\{-\hat{\varepsilon}_{xx}[\Delta\Phi + 2(1-a)\Delta\Psi] + 2\hat{\varepsilon}_{xx}^2\right\}dV = \\ &= 2C_0^2(1-a)G\left\{\iint_V \hat{\varepsilon}_{xx}[F(x)\delta_M + 2(1-a)\hat{\varepsilon}_{xx}]dV - 2\iint_V \hat{\varepsilon}_{xx}^2 dV\right\} = \\ &= 2C_0^2(1-a)G\int_{-L}^{L}\hat{\varepsilon}_{xx}(x,0)F(x)dx - 2a\iint_V \hat{\varepsilon}_{xx}^2 dV = \pi C_0^2(1-a)^2 G \frac{L^2}{2}\end{aligned}$$ (6.19)

Taking into account (4.14) and (6.14) we obtain

$$J_2 = \frac{1-2a}{2a}G\iint_V(\varepsilon_{xx}+\varepsilon_{yy})^2 dV = 2C_0^2 a(1-2a)G\iint_V \hat{\varepsilon}_{xx}^2 dV = \pi C_0^2 a(1-2a)G \frac{L^2}{4}$$ (6.20)

The crack energy by (6.5), (6.19) and (6.20) is

$$W = J_1 + J_2 = \frac{2-3a}{(1-a)^2}\frac{(\sigma^{ex})^2}{G}\frac{\pi L^2}{4}$$ (6.21)

The energy of interaction between the external field and the crack field is $W^i = 0$. Really, the deformation of the external field does not depend on the coordinates. During integration in (6.8) its components are removed from the integral, with the remaining integrals only of the first degrees $\varepsilon$. Taking into account (4.12), (4.13) and (4.22) we obtain

$$\int_V \hat{\varepsilon}_{xx} dV = \frac{1}{2\pi}\int_{-\infty}^{\infty}\int_{-\infty}^{\infty}\int_{-L}^{L} f(t)\frac{x-t}{(x-t)^2+y^2}dtdydx = 0$$

Integration in the sense of the basic value yields

$$\int_{-\infty}^{\infty}\frac{x-t}{(x-t)^2+y^2}dx = \frac{1}{2}\ln[(x-t)^2+y^2]\Big|_{-\infty}^{\infty} = 0$$

For $\omega$ we get

$$\int_{-\infty}^{\infty}\omega dx = \frac{1}{2\pi}\int_{-L}^{L}f(t)\int_{-\infty}^{\infty}\frac{(x-t)y^2}{[(x-t)^2+y^2]^2}dx\,dt = \frac{1}{4\pi}\int_{-L}^{L}f(t)\frac{(x-t)^2}{(x-t)^2+y^2}\Big|_{-\infty}^{\infty}dt = 0$$

$$\int_V \varepsilon_{xy} dV = 0$$

because under the integral there is an odd function in $y$.

The total energy consists of the self-energy and the surface energy. The generalized force acting upon the crack, according to (4.21) and (6.11), is written as

$$g(L) = \pi \frac{2-3a}{2(1-a)^2 G}(\sigma^{ex})^2 GL + 2\gamma > 0$$ (6.22)

and it is always larger than zero, because both the terms are positive. The density of the crack energy (6.5) is a quadratic value and it is always positive. Its integral will be a positive value. Therefore, in the infinite specimen of the critical length there is no crack, as it always tends to close due to the elastic and surface energies.

## 7. Compensation Field Energy

The deformation of a specimen with a surface crack is
$$\bar{\varepsilon} = \varepsilon + \varepsilon^c + \varepsilon^{ex}.$$
The elastic energy for the above deformation is written in the form (6.6) and the potential energy – in the form (6.10) with three values added:

1. The compensation field energy (6.5)
$$W^c = \iint_V \frac{1-2a}{a} G(\varepsilon^c_{xx} + \varepsilon^c_{yy})^2 + 2G\left[(\varepsilon^c_{xx})^2 + (\varepsilon^c_{yy})^2 + 2(\varepsilon^c_{xy})^2\right] dxdy$$

2. The energy of the interaction between the crack field and the compensation field
$$W_2 = 2\iint_V \left[\frac{1-2a}{a} G(\varepsilon_{xx} + \varepsilon_{yy})(\varepsilon^c_{xx} + \varepsilon^c_{yy}) + 2G(\varepsilon_{xx}\varepsilon^c_{xx} + \varepsilon_{yy}\varepsilon^c_{yy} + 2\varepsilon_{xy}\varepsilon^c_{xy})\right] dxdy$$

3. The energy of the interaction between the compensation field and the external field
$$W_3 = 2\iint_V \left[\frac{1-2a}{a} G(\varepsilon^c_{xx} + \varepsilon^c_{yy})(\varepsilon^{ex}_{xx} + \varepsilon^{ex}_{yy}) + 2G(\varepsilon^c_{xx}\varepsilon^{ex}_{xx} + \varepsilon^c_{yy}\varepsilon^{ec}_{yy} + 2\varepsilon^c_{xy}\varepsilon^{ex}_{xy})\right] dxdy$$

Let us calculate on the analogy with (6.12):
$$\int_V \left[(\hat{\varepsilon}^c_{xx})^2 + (\hat{\varepsilon}^c_{xy})^2\right] dV = \int_V |\nabla \hat{U}^c_x|^2 dV = \int_S \hat{U}^c_x \frac{\partial \hat{U}^c_x}{\partial n} dS - \int_V \hat{U}^c_x \Delta \hat{U}^c_x dV =$$
$$= -\int_V \hat{U}^c_x D_x [P * \delta_y] dV = \int_{-\infty}^{\infty} P(y) \frac{\partial \hat{U}^c_x}{\partial x}(0,y) dy = \int_{-\infty}^{\infty} P(y) \hat{\varepsilon}^c_{xx}(0,y) dy =$$
$$= -\frac{1}{\pi} \int_{-\infty}^{\infty} \left[\sqrt{L^2+y^2} \ln \frac{\sqrt{L^2+y^2}+L}{\sqrt{L^2+y^2}-L} + 2L\ln \frac{y}{L}\right]\left[\frac{|y|}{\sqrt{L^2+y^2}} - 1 - \frac{L^2|y|}{(L^2+y^2)^{3/2}}\right] dy = \qquad (7.1)$$
$$= -\frac{2L^2}{\pi} \int_0^{\infty}\left[\sqrt{1+\xi^2} \ln \frac{\sqrt{1+\xi^2}+1}{\sqrt{1+\xi^2}-1} + 2\ln\xi\right]\left[\frac{|\xi|}{\sqrt{1+\xi^2}} - 1 - \frac{|\xi|}{(1+\xi^2)^{3/2}}\right] d\xi = 4{,}27 L^2$$

After proceeding to the dimensionless variables $\xi = y/L$ the integral does not depend on any physical parameters and its numerical value is
$$\frac{2}{\pi}\int_0^{\infty}\left[\sqrt{1+\xi^2} \ln \frac{\sqrt{1+\xi^2}+1}{\sqrt{1+\xi^2}-1} + 2\ln\xi\right]\left[1 - \frac{|\xi|}{\sqrt{1+\xi^2}} + \frac{|\xi|}{(1+\xi^2)^{3/2}}\right] d\xi \approx 4{,}27$$

Since
$$1 - \frac{|\xi|}{\sqrt{1+\xi^2}} + \frac{|\xi|}{(1+\xi^2)^{3/2}} \sim \frac{1}{\xi^2}, \quad \xi \to \infty$$
then the integrals in (7.1) are absolutely converging. Similarly (6.13):
$$\int_V \left[(\hat{\varepsilon}^c_{xy})^2 - (\hat{\varepsilon}^c_{yy})^2\right] dV = 0$$

Hence (7.1) as well as for (6.14):
$$\int_V (\hat{\varepsilon}^c_{xx}) dV = \int_V (\hat{\varepsilon}^c_{yy}) dV = \int_V (\hat{\varepsilon}^c_{xy}) dV = 2{,}13 L^2$$

The compensation field energy is positive and similarly to (6.21) is written as
$$W^c = 4{,}27 \frac{2-3a}{(1-a)^2} \frac{(\sigma^{ex})^2}{G} L^2 \qquad (7.2)$$

## 8. The Energy of Interaction between the Self-Field of the Crack and the Compensation Field

$$W_2 = 2\iint_V \left[\frac{1-2a}{a} G(\varepsilon_{xx} + \varepsilon_{yy})(\varepsilon^c_{xx} + \varepsilon^c_{yy}) + 2G(\varepsilon_{xx}\varepsilon^c_{xx} + \varepsilon_{yy}\varepsilon^c_{yy} + 2\varepsilon_{xy}\varepsilon^c_{xy})\right]dxdy =$$

$$= 2\iint_V \left[\frac{1-2a}{a} G\hat{\varepsilon}_{xx}\hat{\varepsilon}^c_{yy} + 2G(\nabla U_x \nabla U^c_x + \nabla U_y \nabla U^c_y)\right]dxdy$$

We have

$$\iint_V (\hat{\varepsilon}_{xx}\hat{\varepsilon}^c_{xx} + \hat{\varepsilon}_{xy}\hat{\varepsilon}^c_{xy})dV = \iint_V \nabla\hat{U}_x \nabla\hat{U}^c_x dV = \int_S \hat{U}_x \frac{\partial \hat{U}^c_x}{\partial n}dS - \iint_V \hat{U}_x \Delta\hat{U}^c_x dV =$$

$$= -\iint_V \hat{U}_x \Delta\hat{U}^c_x dV = \iint_V \hat{\varepsilon}_{xx} P(y)\delta_y dV = \int_{-\infty}^{\infty} \hat{\varepsilon}_{xx}(0,y)P(y)dy =$$

$$= \frac{L^2}{\pi}\int_0^\infty \left(\frac{\xi}{\sqrt{1+\xi^2}} - 1\right)\left[\sqrt{1+\xi^2}\ln\frac{\sqrt{1+\xi^2}+1}{\sqrt{1+\xi^2}-1} + 2\ln\xi\right]d\xi = -0{,}908L^2$$

As we cannot represent the expression $\hat{\varepsilon}_{xx}\hat{\varepsilon}^c_{xx} - \hat{\varepsilon}_{xy}\hat{\varepsilon}^c_{xy}$ in the form of a scalar product of two gradients like in (6.13), therefore it is necessary to evaluate the integral

$$\iint_V (\hat{\varepsilon}_{xx}\hat{\varepsilon}^c_{xx} - \hat{\varepsilon}_{xy}\hat{\varepsilon}^c_{xy})dV = \frac{1}{4\pi^2}\int_{-\infty}^{\infty}\int_{-L}^{L}\int_{-\infty}^{\infty}\int_{-\infty}^{\infty} f(t)p(t_1)\frac{(x-t)(x-t_1)-y^2}{[(x-t)^2+y^2][(x-t_1)^2+y^2]}dydxdtdt_1$$

First, we integrate with respect to $y$

$$\int_{-\infty}^{\infty}\frac{(x-t)(x-t_1)-y^2}{[(x-t)^2+y^2][(x-t_1)^2+y^2]}dy = \left.\frac{\arctan\frac{y}{t_1-x} - \arctan\frac{y}{t-x}}{t-t_1}\right|_{-\infty}^{\infty} =$$

$$= \pi\frac{\text{sign}(t_1-x) - \text{sign}(t-x)}{t-t_1}$$

Then, with respect to $x$:

$$\int_{-\infty}^{\infty}[\text{sign}(t_1-x) - \text{sign}(t-x)]dx = -2(t_1-t)$$

As a result, we obtain

$$\iint_V (\hat{\varepsilon}_{xx}\hat{\varepsilon}^c_{xx} - \hat{\varepsilon}_{xy}\hat{\varepsilon}^c_{xy})dV = -\frac{1}{2\pi}\int_{-\infty}^{\infty}\int_{-L}^{L} f(t)p(t_1)dtdt_1 = 0$$

because under the integral there is an odd function.

As in the case of (6.14), we obtain

$$\iint_V \hat{\varepsilon}_{xx}\hat{\varepsilon}^c_{xx}dV = \iint_V \hat{\varepsilon}_{yy}\hat{\varepsilon}^c_{yy}dV = \iint_V \hat{\varepsilon}_{xy}\hat{\varepsilon}^c_{xy}dV = -0{,}454L^2$$

And the energy of the interaction between the crack field and the compensation field is written as

$$W_2 = 4(5a+2)G\left[\frac{\sigma^{ex}}{(1-a)G}\right]^2 \int_V \hat{\varepsilon}_{xx}\hat{\varepsilon}^c_{xx}dV = -1{,}82\frac{(5a+2)L^2}{G(1-a)^2}(\sigma^{ex})^2 \qquad (8.1)$$

## 9. The Energy of the Interaction between the External Field and the Compensation Field

The energy of the interaction between the external field and the compensation field is written as

$$W_3 = 2\iint_V \left[ \frac{1-2a}{a} G(\varepsilon_{xx}^c + \varepsilon_{yy}^c)(\varepsilon_{xx}^{ex} + \varepsilon_{yy}^{ex}) + 2G(\varepsilon_{xx}^c \varepsilon_{xx}^{ex} + \varepsilon_{yy}^c \varepsilon_{yy}^{ec} + 2\varepsilon_{xy}^c \varepsilon_{xy}^{ex}) \right] dxdy =$$

$$= 2\frac{(\sigma^{ex})^2}{G} \iint_V \left[ \frac{-4a^2 + a + 1}{(4a-3)(a-1)} \hat{\varepsilon}_{yy}^c + \frac{2}{4a-3} \omega^c \right] dV = \frac{(\sigma^{ex})^2}{G} \frac{4}{4a-3} \iint_V \omega^c dV$$

(9.1)

The integral in (9.1) diverges, that is, $W_3$ is not limited for an infinite specimen and one should use the procedure of cutting and calculate the energy of a specimen of a finite size. Let us evaluate the energy of interaction for a specimen of the rectangular shape and the size along the axis of ordinates from $-A$ to $A$ and along the abcissa axis from $-B$ to $B$. The energy turns out a negative value, i.e. due to it the crack tends to open. Since $W_3 \to -\infty$ at $A, B \to \infty$, then for a crack emerging on the surface of a semi-infinite specimen there is no critical size and it will always open. A critical condition can exist only for a specimen of a finite size. To calculate it accurately, one should solve the problem of a crack in a specimen of a finite size. However, the relation obtained here can be used for evaluation.

We have

$$\iint_V \hat{\varepsilon}_{yy}^c dxdy = \int_{-\infty}^{\infty}\int_{-\infty}^{\infty}\int_{-\infty}^{\infty} p(t) \frac{y-t}{x^2 + (y-t)^2} dydxdt = 0$$

because

$$\int_{-\infty}^{\infty} \frac{y-t}{x^2 + (y-t)^2} dy = \ln\sqrt{x^2 + (y-t)^2} \Big|_{-\infty}^{\infty} = 0$$

For $\omega^c$ taking indefinite integrals with respect to $x$ and $y$ we obtain

$$J = \iint \omega^c dydx = \frac{1}{2\pi} \int_{-A}^{A}\int_{-B}^{B}\int_{-A}^{A} p(t) \frac{(y-t)x^2}{[x^2 + (y-t)^2]^2} dydxdt =$$

$$= \frac{1}{4\pi} \int_{-A}^{A} p(t)(y-t)\arctan\frac{x}{y-t}\Big|_{x=-B}^{x=B} \Big|_{y=-A}^{y=A} dt =$$

(9.2)

$$= \frac{1}{\pi^2} \int_{-A}^{A} \frac{t}{\sqrt{L^2 + t^2}} \operatorname{atanh}\frac{L}{\sqrt{L^2 + t^2}} \left[ (A-t)\arctan\frac{B}{A-t} - (A+t)\arctan\frac{B}{A+t} \right] dt$$

Differentiating we get

$$j = \frac{\partial J}{\partial(2L)} = \frac{1}{\pi^2} \int_0^A \frac{t}{(L^2 + t^2)} \left[ -\frac{L}{\sqrt{L^2 + t^2}} \operatorname{atanh}\frac{L}{\sqrt{L^2 + t^2}} + 1 \right] \cdot$$

$$\cdot \left[ (A-t)\arctan\frac{B}{A-t} - (A+t)\arctan\frac{B}{A+t} \right] dt$$

(9.3)

Under the integral there is an even in $t$ function therefore, integration can be performed from 0 to $A$. The table shows the numerical calculation data (9.3). In the forth column one can see the interval of change of the crack length $L$.

| A | B | j | L |
|---|---|---|---|
| 10000 | 10000 | -340 | 0,1 - 100 |
| 10000 | 1000 | -4,2 | 0,1 - 100 |
| 10000 | 100 | -0,04 | 0,1 - 10 |
| 10000 | 10 | -0,0004 | 0,1 - 1 |
| 1000 | 1000 | -31 | 0,1 - 100 |
| 1000 | 100 | -0,42 | 0,1 - 100 |
| 1000 | 10 | -0,0004 | 0,1 - 1 |
| 100 | 100 | -3,1 | 0,1 - 10 |
| 100 | 10 | -0,31 | 0,1 - 1 |

Note that at fixed values of $A$ and $B$ the value of $j$ is constant accurate to one per cent for a wide range of values of $L$. If $A$ and $B$ simultaneously change $n$-times, $j$ changes correspondingly. Adding (6.21), (7.2), (8.1) and the crack surface energy, differentiating them with respect to $L$ and adding (9.3) we obtain a generalized force acting upon the crack in the form

$$\frac{\partial(W + W^c + W_2 + 4\gamma)}{\partial(2L)} + j = 0{,}965 \frac{(5a+2)}{(1-a)^2} \frac{(\sigma^{ex})^2 L}{G} + 2\frac{(\sigma^{ex})^2}{G} j + 2\gamma \qquad (9.4)$$

Equating (9.4) to zero we derive a critical stress

$$\sigma_*^{ex} = \sqrt{\frac{\gamma G}{-j - 0{,}48 \frac{(5a+2)}{(1-a)^2} L}} \qquad (9.5)$$

The denominator in (9.5) will be larger than zero for a specimen of sufficiently large size, since $j < 0$. (9.5) yields the dependence of the critical stress on the crack length. Besides, when the specimen thickness decreases, that is, the size of $B$, then

$$-j \to 0{,}48 \frac{(5a+2)}{(1-a)^2} L$$

Hence $\sigma_*^{ex} \to \infty$. In this case, the theoretical strength of the material $\sim G$ is achieved. It is a well-known experimental fact.

Deformation for a flat crack in the infinite specimen was obtained by Griffith [6,9]. He also calculated approximately its elastic energy. For some unexplained reason, he equated the above energy to the surface energy and obtained (9.5). As a result, he derived the critical condition for a crack in the infinite specimen. Such equating is meaningless in physics.
Apparently, he had to do it to get a theoretical basis for the experimentally determined relation

$$\sigma_*^{ex} \sim 1/\sqrt{L} \qquad (9.6)$$

According to the general laws of physics, the elastic and surface energies should be added and the potential energy of the system should be found, as it was done in (6.22). Then we get a result contradicting that obtained by Griffith and, namely, there is no critical stress for an infinite specimen, which is clear without any calculations. Both the elastic and surface energies are positive and increase with growing crack. Therefore, they can only close crack. Irwin introduced the notion of a stress concentrator [7] as a coefficient at $1/\sqrt{L}$ in the stress field near the crack head. It is postulated that for each material there exists a critical value of the stress concentrator above which the crack grows spontaneously. The possibility of the crack growth due to the elastic energy was not considered in [7]. The two theories were necessary to explain the experimental fact (9.6). As a matter of fact, the reason for the critical condition of the crack in a specimen of the finite size is the elastic energy of the interaction between the external stress and the field of stress caused by the deformation of the free surface under the action of the crack.

The critical condition is characteristic not only of the crack emerging on the free surface but also of the crack located nearly at a distance of several lengths of the crack.

### 10. Discussion

The technique of solution of Poisson's equation suggested here has been illustrated by the example of the two-dimensional theory of elasticity. In [10] a previously unknown solution of the electron problem was found, that is, the equation (1.21) was solved in the three-dimensional case. The equation (1.21) called a continuity equation, enters the system of equations of continuum mechanics and is used in the gravitation theory, the theory of electricity, etc. The same is true for Poisson's equation. The results obtained here can be useful for the solution of the problems arising in various fields of physics, where these equations occur.